\documentstyle[11pt,paspconf,epsf]{article}

\begin{document}

\title{Multiscaling}
\author{Vicent J. Mart\'{\i}nez}

\affil{Departament d'Astronomia i
Astrof\'{\i}sica, Universitat de Val\`encia,
E-46100 Burjassot, Val\`encia, Spain}

\author{Mar\'{\i}a Jes\'us Pons-Border\'{\i}a\altaffilmark{1}}
\affil{Departamento de F\'{\i}sica Te\'orica, Universidad
Aut\'onoma de Madrid, E-28049 Madrid, Cantoblanco, Spain}

\altaffiltext{1}{Departament de Matem\`atica Aplicada i Astronomia,
Universitat de Val\`encia,
E-46100 Burjassot, Val\`encia, Spain}

\begin{abstract}
We introduce the unbiased way statisticians look at the 2--point
correlation function and study its relation to multifractal analysis.
We apply this method to a simulation of the distribution of galaxy clusters
in order to check the dependence of the correlation dimension on the cluster
richness.
\end{abstract}

\section{Introduction}

The statistical description of the galaxy clustering is usually
based on the two-point correlation function $\xi(r)$. This
function is, following the terminology used by statisticians
working in point field statistics, a second-order characteristic
of the point process (Diggle 1993; Stoyan \& Stoyan 1994). The first-order
characteristic is just the intensity measure $\lambda(\vec r)$
(Mart\'{\i}nez et al. 1993). Assuming the
Cosmological principle, we accept that galaxies in large volumes
represent a stationary and isotropic point
process, having therefore constant intensity equal to the number density
of galaxies per unit volume, denoted by $n$.

\section{The $K$-function and the correlation dimension}
Among all the second-order characteristics of a point-process the
most commonly used is the Ripley's $K$-function (Ripley 1981)
which is easily
related to the mean number of galaxies lying in balls of radius $r$
centred in an arbitrary galaxy, $\langle N \rangle_r$
(Peebles 1980), or the correlation integral $C(r)$ (Mart\'{\i}nez et al
1995). For a three-dimensional process we have that:
\begin{equation}
\langle N \rangle_r  = n K(r) = C(r) = \int_0^r 4 \pi n s^2 (1 +
\xi(s)) ds
\end{equation}

It is obvious that for a homogeneous Poisson process the
$K$-function is
\begin{equation}
K(r) = {4 \pi \over 3 } r^3
\end{equation}

It is well known the power-law behaviour of $\xi(r)$ for the
galaxy distribution, $\xi(r)=(r/r_0)^{-\gamma}$ with
$\gamma \simeq 1.8$ and $r_0 \simeq 5 \, h^{-1}$ Mpc (Davis \& Peebles
1983). In the strong
clustering regime where $\xi(r) \gg 1$, the previous behaviour
is translated to a power-law behaviour in the Ripley's function:
$K(r) \propto r^{D_2}$ with $D_2 \simeq 3 -\gamma$. The exponent
$D_2$ is known as the correlation dimension. We have argued that
this value itself is an important measure of the clustering
(Mart\'{\i}nez et al. 1995).
Several authors have successfully fitted a power-law
to the cluster-cluster correlation function.
Some of them claimed to have obtained a value of
1.8 for $\gamma$ (Bahcall \& Soneira
1983), but many others have not.
Recent
calculations give no hope for a common value of the
exponent $\gamma$ for clusters and galaxies (Postman et al. 1992).
For example Dalton et al. (1994) found that $\gamma \simeq 2$ for
the APM clusters.
The correlation length found for clusters is larger than
the one for galaxies, $r_{0,c} \simeq
15 - 30 \, h^{-1}$ Mpc. However a strong
controversy has arosen about the value of $r_{0,c}$
(Bahcall \& Soneira 1983; Bahcall et al. 1986, Sutherland 1988;
Sutherland \& Efstathiou, 1991).
In this context one has to be cautious in
interpreting the power-law behaviour of $\xi(r)$ for clusters.
Moreover, in the range of scales where $\xi(r)$ is of the order of
unity it is obvious that $K(r)$ and $\xi(r)$ cannot simultaneously
display power-law behaviour. This could be the case for the
distribution of clusters, which in the range [10, 50] $h^{-1}$ Mpc
present values less than 5 for $\xi(r)$. We have seen that probably
$K(r)$ fits better a power-law in this range than $\xi(r)$ does and
the exponent $D_2$ is different for galaxies and for clusters
(Mart\'{\i}nez et al. 1995).

\section{Multiscaling}
This variation of the relevant exponents with the
the kind of object, galaxies of different morphological type or
clusters with different richness, might be considered within the
framework of the multiscaling hypothesis (Jensen et al. 1991, Paladin
et al. 1992). If all these
astrophysical objects are interpreted as the consequence of
applying different mass thresholds to a continuous density field,
the multiscaling approach provides us with a theoretical way to
understand the variations in the clustering properties which turn out
in variations of the exponent $D_2$. The higher the density
threshold the stronger the clustering and therefore the smaller
the value of the corresponding $D_2$. We have tested this
approach on several cluster and galaxy simulated samples and we conclude
that the multiscaling approach is rather useful for the
description of the large scale clustering in the Universe.
In fact the dependence of the clustering properties of clusters with
different richness is a natural example to illustrate how
multiscaling works (Paredes et al. 1995).
We shall see here that the use of the function
$K(r)$ with appropriate unbiased estimators is the best tool
for the study of the multiscaling distribution in models of clusters of
galaxies and at the same time gives us information about the scale
where the transition to homogeneity appears.
\section{Estimators}

The estimators of
a statistical quantity as $K(r)$ need to have several good
properties (Stoyan \& Stoyan 1994), but obviously the most important one
is that the estimator
must be unbiased, in the sense that the mean value of the
estimates when applied to different samples of a given process
has to be equal to the true value of the statistical quantity.
At large distances, the unbiasedness of $K(r)$ is mostly related
with the edge-corrections of the estimators. Here we shall use the
estimator proposed by Ripley (1976) and
Baddeley et al. (1993). For $N$ galaxies at
positions $\{X_i\}_{i=1}^N$ distributed in a volume $V$, it reads:
\begin{equation}
\hat K (r) = {V\over N^2} \sum_{i=1}^N \sum^N_{\scriptstyle
j=1 \atop\scriptstyle j\ne i }
{\theta(r-|{\vec X}_i -{\vec X}_j|)\over w_{ij}},
\end{equation}
$\theta$ being the step function, which is 0 for negative values of
its argument and 1 for positive ones. The weight $w_{ij}$ is the
proportion of the area of the sphere, centred at $X_i$ and passing
through $X_j$, which lies inside the volume $V$. In the case of a cubic
sample such as we consider, Baddeley et al. (1993) give a (complicated)
analytic expression for it that we can use. The sum in Eq. (3) is an
unbiased estimator for $n^2 V K(r)$ (Ripley 1976, 1981; Diggle 1983).
For more details see also Mart\'{\i}nez \& Pons (1996).

\section{Results in a model for the distribution of clusters}
The fractal behaviour of the distribution of galaxy clusters has
a breakdown at some given distance (Borgani et al. 1994), in which the
tendency to homogeneity appears in an evident way.
Here we shall use a numerical model for the
distribution of rich clusters. In this model, clusters are
identified as the peaks in the evolved density field of Zel'dovich
simulations with $\Omega = 1$, where 30\% of this critical density
is provided by massive neutrinos and the rest by cold particles
(Klypin et al. 1993, Borgani et al. 1995).
Applying different density thresholds to the
simulation we obtain samples ressembling the samples of rich
galaxy clusters. The higher the threshold the richer the corresponding
clusters and the larger the mean separation between them. Using the
mean particle separation $d = n^{-1/3}$ as the control parameter
we have extracted 5 samples from one realization of the model.
In Figure 1, we show these samples drawn from a simulation in a cube
of side $320 \, h^{-1}$ Mpc.

\begin{figure}
\plotone{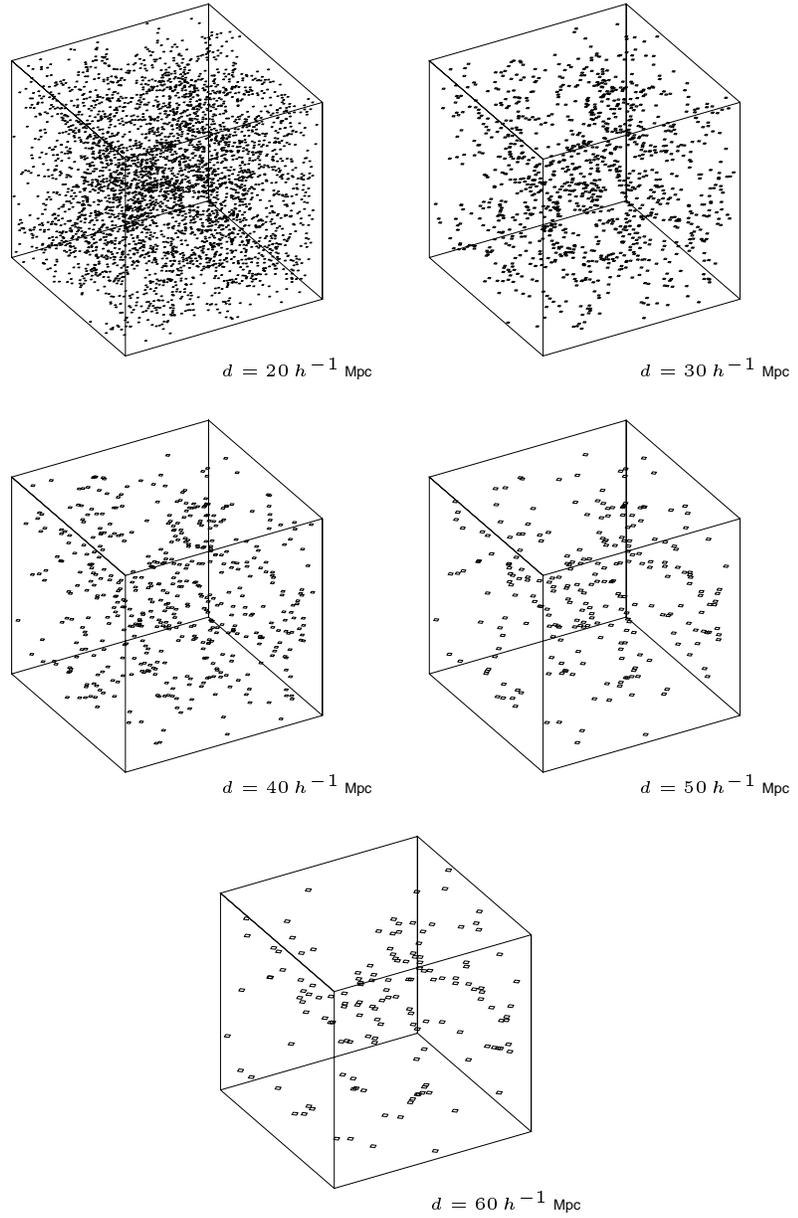}
\caption{Cluster samples drawn from the simulations in a cubic volume
of side 320 $h^{-1}$ Mpc. The mean interparticle distance $d$ is
indicated in each case.}
\end{figure}

In Figure 2 we show the $K$-function for each of the previous samples,
together with the variation of the correlation dimension $D_2$
with the scale $r$.
This latter value has been calculated as the local slope of the log-log
plot $K(r)$ vs. $r$ in a range of scales of 10 $h^{-1}$ Mpc width
centred at each value of $r$.
We can see that for scales in the range [10, 40] $h^{-1}$
Mpc the function $K(r)$ follows reasonably well a power law with exponent
$D_2 < 3$. However the exponent depends on the density in the way prescribed
by the multiscaling hypothesis: for higher density threshold
(or equivalently larger
interparticle distances), smaller values of $D_2$ are found, indicating
stronger clustering. For scales larger than 40 $h^{-1}$ Mpc, a clear tendency
to homogeneity (represented by the dotted line in the lower panel) is
observed. In the plots of $D_2$ as a function of the scale, we can
clearly see these two kinds of behaviour. First a noticeable plateau,
in which the value of $D_2$ oscillates without any clear trend around
a constant value for each sample.
This happens for small scales. For larger scales the local value of $D_2$
increases with $r$, indicating a clear tendency to homogeneity, which is
formally reached around  100 $h^{-1}$ Mpc, where $D_2 \simeq 3$.

\begin{figure}[h]
\plotone{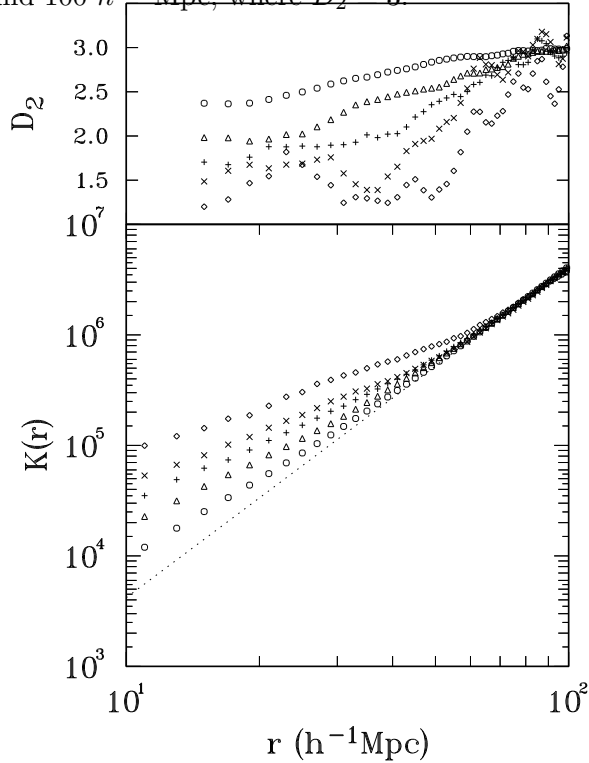}
\caption{In the bottom panel we see the $K$-function for the
cluster samples shown in Figure 1. The
dotted line corresponds to the expected theoretical function of a Poisson
distribution (slope = 3).
The different point marks correspond from bottom to top
to progressively higher values of the mean interparticle separation.
The top panel shows the variation of the local slope $D_2$ with the scale.}
\end{figure}


\acknowledgements
We are grateful to S. Borgani, P. Coles, L. Moscardini and M. Plionis
for permission to use cluster simulations done by them and for useful
discussions. We thank R. Moyeed for his kind information on properties
of $K$. This work has been  supported by the {\em Conselleria
d'Educaci\'o i Ci\`encia de la Generalitat Valenciana} (grant number
GV-2207/94)
and by the {\it Instituci\'o Valenciana d'Estudis i Investigaci\'o}.

\end{document}